\documentclass[sigconf]{acmart}
\AtBeginDocument{%
  }

\usepackage{hyperref}       
\usepackage{amsfonts}       
\usepackage{bm}
\usepackage{graphicx}
\usepackage{amsthm}
\usepackage{amsmath}
\usepackage{multirow}
\usepackage{enumitem}
\usepackage{bbding}
\usepackage{colortbl}
\usepackage{arydshln}
\usepackage{algorithm}  
\usepackage{algorithmicx}  
\usepackage{threeparttable}
\usepackage{algpseudocode}  
\usepackage{subfigure}

\copyrightyear{2025}
\acmYear{2025}
\setcopyright{acmlicensed}
\acmConference[CIKM '25] {Proceedings of the 34th ACM International Conference on Information and Knowledge Management}{ November 10--14, 2025}{Seoul, Republic of Korea.}
\acmBooktitle{Proceedings of the 34th ACM International Conference on Information and Knowledge Management (CIKM '25), November 10--14, 2025, Seoul, Republic of Korea}
\acmISBN{979-8-4007-2040-6/2025/11}
\acmDOI{10.1145/XXXXXX.XXXXXX}

\begin{document}

\title{PrLM: Learning Explicit Reasoning for Personalized RAG via Contrastive Reward Optimization
}

\author{Kepu Zhang\textsuperscript{{$\star$}}}
\author{Teng Shi\textsuperscript{{$\star$}}}
\affiliation{
\institution{\mbox{Gaoling School of Artificial Intelligence}\\Renmin University of China}
  \city{Beijing}
  \country{China}
}
\email{{kepuzhang,shiteng}@ruc.edu.cn}

\author{Weijie Yu}
\authornote{Weijie Yu is the corresponding author. 
\\ {$\star$}  Equal contribution. Work partially done at Beijing Key Laboratory of Research on Large Models and Intelligent Governance, and Engineering Research Center of Next-Generation Intelligent Search and Recommendation, MOE.
}
\affiliation{
\institution{School of Information Technology and Management\\University of International Business and Economics}
  \city{Beijing}
  \country{China}
}
\email{yu@uibe.edu.cn}

\author{Jun Xu}
\affiliation{
\institution{\mbox{Gaoling School of Artificial Intelligence}\\Renmin University of China}
  \city{Beijing}
  \country{China}
}
\email{junxu@ruc.edu.cn}

\renewcommand{\shortauthors}{Kepu Zhang et al.}

\begin{abstract}
Personalized retrieval-augmented generation (RAG) aims to produce user-tailored responses by incorporating retrieved user profiles alongside the input query.
Existing methods primarily focus on improving retrieval and rely on large language models (LLMs) to implicitly integrate the retrieved context with the query. However, such models are often sensitive to retrieval quality and may generate responses that are misaligned with user preferences.
To address this limitation, we propose \texttt{PrLM}, a reinforcement learning framework that trains LLMs to explicitly reason over retrieved user profiles. Guided by a contrastively trained personalization reward model, \texttt{PrLM} effectively learns from user responses without requiring annotated reasoning paths.
Experiments on three personalized text generation datasets show that \texttt{PrLM} outperforms existing methods and remains robust across varying numbers of retrieved profiles and different retrievers.

\end{abstract}




\maketitle

\section{Introduction}\label{sec:intro}
Personalized retrieval-augmented generation (RAG)~\cite{salemi2024lamp,salemi2024optimization} aims to generate user-specific responses by incorporating retrieved user profiles alongside the input query. This task has drawn increasing attention due to its broad applicability in personalized dialogue, summarization, recommendation explanation, and beyond~\cite{shen2025stylitruth}. By combining the strengths of retrieval and generation~\cite{jiang2023active,zhang2025qe,zhang2024citalaw}, personalized RAG systems can provide responses that are both grounded in external knowledge and tailored to individual users.

\begin{figure}[t]
    \centering
    \includegraphics[width=\linewidth]{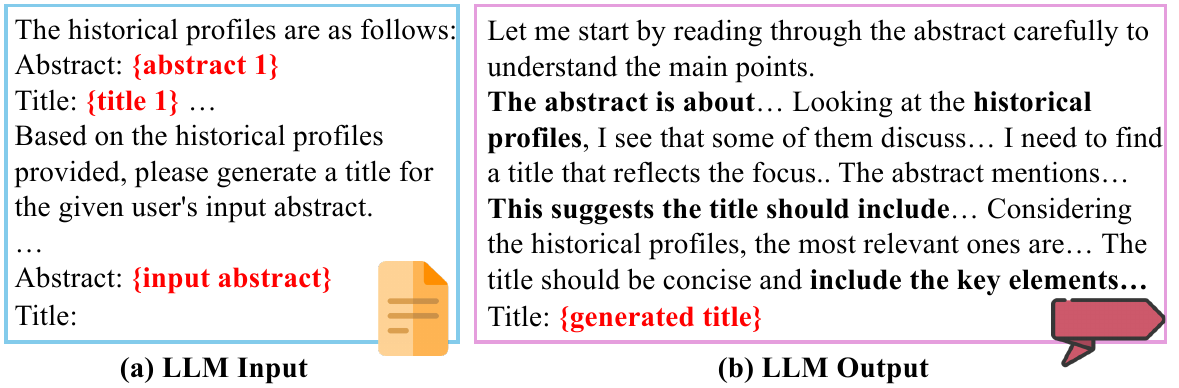}
    \caption{
     An example of explicit reasoning in personalized RAG from the LaMP-5 dataset.
(a) The input includes the user’s historical abstracts and titles, along with the target abstract.
(b) The LLM trained with \texttt{PrLM} generates a personalized title by explicitly reasoning over the retrieved user profile and the target abstract.
    }
    \label{fig:fig1}
    \vspace{-3mm}
\end{figure}

Recent studies~\cite{richardson2023integrating,chen2024large,shi2025retrieval} in personalized RAG have primarily focused on enhancing the retrieval stage. These methods aim to select the most relevant user profiles and concatenate them with the query before inputting the combined context into a large language model (LLM). The LLM is then expected to implicitly integrate the retrieved information and produce a personalized response. 
Despite achieving promising results, this implicit reasoning paradigm is highly sensitive to retrieval noise. 
Motivated by recent advances in LLM reasoning~\cite{xu2025towards,chen2025towards}, we argue that introducing an explicit reasoning process—where the LLM first reflects on the retrieved context before producing the final output—has the potential to improve the robustness, and personalization fidelity.
Figure~\ref{fig:fig1} illustrates an example of explicit reasoning in a personalized RAG. The LLM analyzes the target abstract in light of the user’s historical profiles and generates an intermediate reasoning trace that guides the final personalized title. This structured process highlights the model’s capacity to align more faithfully with user preferences.

However, implementing explicit reasoning in personalized RAG poses two major challenges.
\textbf{First}, there is a lack of annotated intermediate reasoning paths, making it difficult to directly supervise the reasoning process during training. 
\textbf{Second}, current training objectives do not account for the degree of personalization in the generated responses. To effectively guide explicit reasoning, it is necessary to design a reward model that can evaluate and optimize for personalization quality.

To address the above challenges, we propose \texttt{PrLM}, a reinforcement learning framework that enables LLMs to explicitly reason over retrieved user profiles in personalized RAG. We leverage GRPO~\cite{shao2024deepseekmath} to train the LLM by sampling and optimizing over groups of reasoning paths without requiring intermediate supervision. This approach encourages the model to learn a structured decision process that infers user preferences  before generating the final response. 
By modeling this explicit reasoning, \texttt{PrLM} enhances the model’s ability to effectively leverage personalized context and improves the faithfulness of the generated outputs. 
Moreover, to assess the degree of personalization, we further introduce a contrastively trained reward model. For the same input query, we compare the LLM’s responses generated with and without retrieved user profiles, and train the reward model to assign higher scores to responses that better reflect user-specific information. This personalization reward is then integrated into the GRPO framework to guide learning.
Experiments on three personalized text generation datasets demonstrate that \texttt{PrLM} consistently outperforms existing baselines and maintains robustness across varying numbers of retrieved profiles and different retrievers.

To summarize, our contributions are as follows:
\begin{itemize}[leftmargin=*]
    \item We propose \texttt{PrLM}, a novel reinforcement learning framework that enables large language models to perform explicit reasoning over retrieved user profiles for personalized RAG.
    
    \item We introduce a contrastive personalization reward model that provides supervision for personalization by comparing outputs with and without retrieved profiles, guiding LLMs to favor more personalized outputs.
    
    \item Extensive experiments on three personalized RAG datasets show \texttt{PrLM} improves both personalization quality and robustness across different retrieval settings and profile quantities.
    
\end{itemize}

\section{Related Work}
\subsection{Personalized RAG}
In recent years, personalization has gradually drawn people's attention~\cite{shi2024unisar,zhang2024saqrec,shi2025unified}.
The personalized RAG task~\cite{salemi2024lamp,salemi2024optimization,mysore2023pearl}, based on a user's historical behavior, can generate personalized outputs for the same query according to different users. 
Existing personalized RAG methods~\cite{shi2025retrieval,li2023teach,zhuanghydra} mainly focus on how to integrate the information of user historical behavior into the prompt of the LLM, so that the LLM can implicitly summarize the user's preferences. 
In this paper, we mainly explore the integration of reasoning with personalized RAG and investigate whether the proposed method can be adapted to various retrieval methods.

\subsection{LLM Reasoning}
Recently, reasoning LLMs have emerged with the advent of models like Deepseek-R1~\cite{guo2025deepseek}. 
Numerous studies have begun to explore various aspects of enhancing the reasoning capabilities of LLMs~\cite{ye2025limo,aggarwal2025l1,zhang2025legal,zhang2024beyond} and leveraging these capabilities to solve a wide range of problems~\cite{xie2025logic,lin2025rec,chen2024huatuogpt,li2025system,shi2025benefit,shi2025bridging,zhang2025syler}.
In this paper, we primarily investigate reasoning in the context of personalized RAG. We explore how to implement reasoning in personalized RAG and utilize reasoning to help LLMs achieve better personalized outputs.

\section{Method}\label{sec:method}

\begin{figure}
    \centering
    \includegraphics[width=0.45\textwidth]{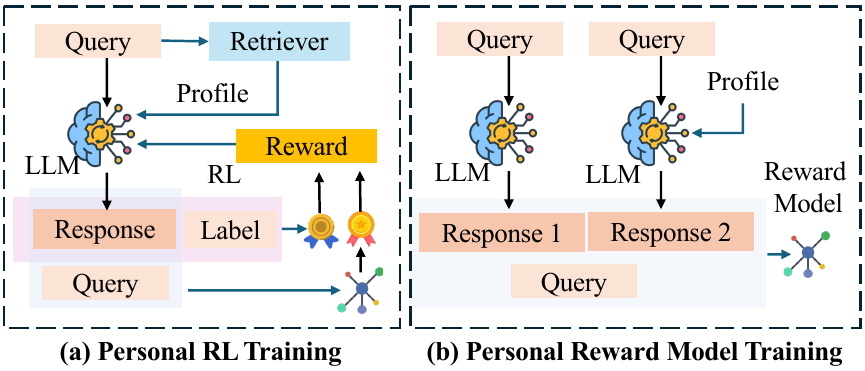}
    \caption{ The architecture of PrLM.
    }
    \label{fig:framework}
    \vspace{-3mm}
\end{figure}

\begin{table}[t]
\centering
        \setlength{\abovecaptionskip}{0pt} 
    \setlength{\belowcaptionskip}{0pt} 
\caption{Statistics of used datasets. 
}
\resizebox{0.7\columnwidth}{!}{
\begin{tabular}{lccccc}
    \toprule
    \multirow{1}{*}{\textbf{Dataset}} &\multirow{1}{*}{\#\textbf{User}} &\multirow{1}{*}{\#\textbf{Train}} &\multirow{1}{*}{\textbf{Dev}} 
    &\multirow{1}{*}{\textbf{Test}}\\
    \midrule
    \textbf{LaMP-4} &1,643&12,500&1,500&1,800\\
    \textbf{LaMP-5} &14,682&14,682&1,500&1,500\\
    \textbf{LaMP-7} &13,437&13,437&1,498&1,500\\
    \bottomrule
\end{tabular}}
\label{tab:data_main}
\vspace{-3mm}
\end{table}

\begin{table*}[t]
    \centering
    \caption{The main experimental results on the LaMP-4, LaMP-5, and LaMP-7 datasets. Bold indicates the best results.
    }
    \resizebox{0.8\textwidth}{!}{
    \begin{tabular}{l|cccc|cccc|cccc}
        \toprule
          &\multicolumn{4}{c|}{\textbf{LaMP-4}} & \multicolumn{4}{c|}{\textbf{LaMP-5}} & \multicolumn{4}{c}{\textbf{LaMP-7}} \\
        \cmidrule(lr){2-5} \cmidrule(lr){6-9} \cmidrule(lr){10-13}
          \textbf{Model} & \textbf{Rouge-1} & \textbf{Rouge-2} & \textbf{Rouge-L} & \textbf{BLEU} & \textbf{Rouge-1} & \textbf{Rouge-2} & \textbf{Rouge-L} & \textbf{BLEU} & \textbf{Rouge-1} & \textbf{Rouge-2} & \textbf{Rouge-L} & \textbf{BLEU}  \\ 
        \hline
        Zero-Shot & 1.29 & 0.25 & 1.03 & 2.23 & 25.19 & 11.79 & 20.47 & 27.34 & 26.83 & 9.69 & 20.36 & 23.74 \\
Random & 8.21 & 1.73 & 7.11 & 9.97 & 26.10 & 10.99 & 21.35 & 30.39 & 30.85 & 11.89 & 25.00 & 28.61 \\
Recency & 8.36 & 1.96 & 7.29 & 10.64 & 28.69 & 12.65 & 23.67 & 32.65 & 30.35 & 11.79 & 24.85 & 28.52 \\
BM25 & 8.91 & 1.73 & 7.61 & 10.76 & 31.07 & 13.72 & 25.20 & 35.20 & 31.03 & 11.90 & 25.26 & 29.17 \\
BGE & 9.66 & 2.15 & 8.39 & 11.78 & 30.21 & 13.45 & 24.91 & 34.03 & 32.46 & 12.90 & 26.67 & 30.57 \\
ROPG & 9.88 & 2.30 & 8.49 & 11.95 & 31.19 & 13.56 & 25.25 & 34.66 & 31.68 & 12.52 & 25.92 & 29.60 \\
CFRAG & 9.76 & 2.23 & 8.34 & 12.06 & 31.41 & 14.56 & 26.05 & 34.59 & 33.42 & 13.11 & 27.45 & 31.49 \\
\hdashline
PrLM & \textbf{12.02} & 2.59 & \textbf{10.35} & \textbf{14.69} & \textbf{36.74} & \textbf{17.25} & \textbf{30.93} & \textbf{40.56} & \textbf{35.96} & \textbf{14.41} & \textbf{30.11} & \textbf{33.80} \\
w/o $r_\mathrm{personal}$ & 11.03 & \textbf{2.66} & 9.43 & 13.28 & 35.07 & 16.40 & 29.40 & 38.82 & 33.98 & 13.58 & 28.55 & 31.97 \\

        \bottomrule
    \end{tabular}
    }
    \vspace{-2mm}
    \label{tab:merged_results}
\end{table*}

\subsection{Task Formulation}
The goal of personalized RAG is to generate a user-specific response $y$ based on a user query $x$ and relevant historical profile $H$:
\begin{equation}
\label{eq:llm}
y = \text{LLM}(x, H).
\end{equation}
The historical profile $H$ is retrieved from a user-specific corpus $D$, which contains the user’s past behavior records (e.g., previous queries, responses, or interactions) given user query $x$:
\begin{equation}
\label{eq:retrieval}
H = \text{Retrieve}(x, D).
\end{equation}
While prior work largely targets improving the retrieval step—i.e., obtaining higher-quality $H$—our focus lies in the generation process itself. We aim to train the LLM to reason explicitly over retrieved profiles and to produce personalized responses that faithfully reflect user preferences, even in the absence of explicit reasoning supervision. To this end, we introduce a group-based reinforcement learning strategy to guide reasoning (§\ref{sec:rl}) and a contrastive reward model to assess personalization quality (§\ref{sec:prm train}).

\subsection{Personalized Reasoning Model Training}
\label{sec:rl}
To enable the LLM to learn explicit reasoning in the absence of annotated reasoning paths, we adopt Group Relative Policy Optimization (GRPO)~\cite{shao2024deepseekmath}, a reinforcement learning algorithm well-suited for optimizing over multiple sampled reasoning trajectories without relying on a value function. GRPO encourages the model to explore diverse reasoning strategies and preferentially reinforce those that yield higher-quality personalized outputs. Formally, Given a user query $x$ and the retrieved user profile $H$, the LLM generates a response $o$ composed of two parts: an intermediate reasoning path $R$ enclosed in \texttt{<think>...</think>} and a final personalized output $y$ following the reasoning. During training, multiple such outputs are sampled, and GRPO updates the model using their relative ranking based on a composite reward signal. The total reward assigned to a sampled response is defined as:
\begin{equation}
\label{eq:reward}
r = r_\mathrm{correct} + \alpha r_\mathrm{think} + \beta r_\mathrm{personal},
\end{equation}
where $\alpha$ and $\beta$ are tunable hyperparameters controlling the importance of each component. The reward terms are defined as follows:
\begin{itemize}[leftmargin=*]
    \item \textbf{Format reward} $r_\mathrm{think}$: this binary reward verifies whether the generated output includes a \texttt{<think>} segment, encouraging the model to explicitly reason before generating the final response.
    \item \textbf{Correctness reward} $r_\mathrm{correct}$: this reward measures the textual similarity between the generated output $y$ and the annotated reference using the sum of ROUGE-1, ROUGE-2, and ROUGE-L scores. Although it captures basic response quality, it only weakly reflects personalization.
    \item \textbf{Personalization reward} $r_\mathrm{personal}$: this component is computed by a separately trained reward model that quantifies the alignment between the output and the user’s historical preferences. Details are provided in §\ref{sec:prm train}.
\end{itemize}

We initialize the LLM using \texttt{DeepSeek-R1-Distill-1.5B}. To control training cost, we fine-tune the model on a randomly selected subset of 500 training samples.

\subsection{Personalized Reward Model Training}\label{sec:prm train}
To quantitatively evaluate the degree of personalization in LLM outputs, we train a reward model via preference-based learning inspired by Direct Preference Optimization (DPO)~\cite{rafailov2023direct}. Unlike conventional response quality metrics, our reward model is explicitly designed to capture whether the generated output reflects user-specific preferences.

We implement the reward model as a BERT~\cite{devlin2019bert} encoder, taking as input the concatenated pair ``[CLS] $x$ [SEP] $y$ [SEP]'', where $x$ is the user query and $y$ is the generated response. The model outputs a scalar score representing the degree of personalization for the $(x, y)$ pair.
To construct training data, we collect 5{,}000 triplets $(x, y_+, y_-)$, where $y_+$ is a personalized response generated by conditioning the LLM on both the query $x$ and retrieved profile $H$, and $y_-$ is a non-personalized response generated in a zero-shot setting (i.e., using $x$ alone). These paired comparisons allow us to train the reward model to prefer responses that reflect personalized content. The optimization objective is a contrastive preference loss:
\begin{equation}
\label{eq:rm}
\mathcal{L} = -\log\left( \sigma(s_p - s_n) \right),
\end{equation}
where $\sigma(\cdot)$ denotes the sigmoid function, and $s_p$ and $s_n$ are the model's predicted scores for the preferred and non-preferred responses, respectively:
\begin{equation}
\label{eq:score}
s_p = \mathrm{BERT}(x, y_+), \quad s_n = \mathrm{BERT}(x, y_-)
\end{equation}

We use BGE~\cite{bge_embedding} as the retriever to construct the personalized inputs for generating $y_+$. Once trained, the reward model can be applied to score the personalization degree of any output $y$ given a query $x$:
\begin{equation}
\label{eq:rm reward}
r_\mathrm{personal} = \mathrm{BERT}(x, y).
\end{equation}

\section{Experiments}\label{sec:experiments}

\begin{table*}[t]
    \centering
    \caption{The retriever robustness results on the LaMP-4, LaMP-5, and LaMP-7 datasets. Bold indicates the best results.
    }
    \resizebox{0.75\textwidth}{!}{
    \begin{tabular}{l|cccc|cccc|cccc}
        \toprule
          &\multicolumn{4}{c|}{\textbf{LaMP-4}} & \multicolumn{4}{c|}{\textbf{LaMP-5}} & \multicolumn{4}{c}{\textbf{LaMP-7}} \\
        \cmidrule(lr){2-5} \cmidrule(lr){6-9} \cmidrule(lr){10-13}
          \textbf{Model} & \textbf{Rouge-1} & \textbf{Rouge-2} & \textbf{Rouge-L} & \textbf{BLEU} & \textbf{Rouge-1} & \textbf{Rouge-2} & \textbf{Rouge-L} & \textbf{BLEU} & \textbf{Rouge-1} & \textbf{Rouge-2} & \textbf{Rouge-L} & \textbf{BLEU}  \\ 
        \hline
Zero-Shot & 1.29 & 0.25 & 1.03 & 2.23 & 25.19 & 11.79 & 20.47 & 27.34 & 26.83 & 9.69 & 20.36 & 23.74 \\
+PrLM & \textbf{3.21} & \textbf{0.71} & \textbf{2.75} & \textbf{4.19} & \textbf{34.36} & \textbf{17.22} & \textbf{28.80} & \textbf{37.33} & \textbf{34.45} & \textbf{13.37} & \textbf{28.53} & \textbf{33.12} \\
\hdashline
Random & 8.21 & 1.73 & 7.11 & 9.97 & 26.10 & 10.99 & 21.35 & 30.39 & 30.85 & 11.89 & 25.00 & 28.61 \\
+PrLM & \textbf{10.10} & \textbf{1.97} & \textbf{8.68} & \textbf{12.50} & \textbf{33.23} & \textbf{15.22} & \textbf{28.01} & \textbf{37.57} & \textbf{35.62} & \textbf{14.11} & \textbf{29.86} & \textbf{33.27} \\
\hdashline
Recency & 8.36 & 1.96 & 7.29 & 10.64 & 28.69 & 12.65 & 23.67 & 32.65 & 30.35 & 11.79 & 24.85 & 28.52 \\
+PrLM & \textbf{11.17} & \textbf{2.31} & \textbf{9.73} & \textbf{13.78} & \textbf{35.15} & \textbf{16.56} & \textbf{29.70} & \textbf{39.16} & \textbf{36.20} & \textbf{14.55} & \textbf{30.29} & \textbf{34.23} \\
\hdashline
BM25 & 8.91 & 1.73 & 7.61 & 10.76 & 31.07 & 13.72 & 25.20 & 35.20 & 31.03 & 11.90 & 25.26 & 29.17 \\
+PrLM &\textbf{10.95} & \textbf{2.06} & \textbf{9.44} & \textbf{13.43} & \textbf{36.56} & \textbf{17.22} & \textbf{30.55} & \textbf{40.38} & \textbf{36.58} & \textbf{14.52} & \textbf{30.68} & \textbf{34.32} \\
\hdashline
ROPG & 9.88 & 2.30 & 8.49 & 11.95 & 31.19 & 13.56 & 25.25 & 34.66 & 31.68 & 12.52 & 25.92 & 29.60 \\
+PrLM& \textbf{11.92} & \textbf{2.42} & \textbf{10.18} & \textbf{14.51} & \textbf{36.24} & \textbf{16.80} & \textbf{30.21} & \textbf{39.57} & \textbf{36.38} & \textbf{14.55} & \textbf{30.45} & \textbf{34.29} \\
\hdashline
CFRAG & 9.76 & 2.23 & 8.34 & 12.06 & 31.41 & 14.56 & 26.05 & 34.59 & 33.42 & 13.11 & 27.45 & 31.49 \\
+PrLM & \textbf{12.50} & \textbf{2.66} & \textbf{10.68} & \textbf{14.90} & \textbf{37.22} & \textbf{17.68} & \textbf{31.20} & \textbf{40.83} & \textbf{36.54} & \textbf{14.70} & \textbf{30.62} & \textbf{34.32}  \\
        \bottomrule
    \end{tabular}
    }
    \vspace{-2mm}
    \label{tab:retrieve}
\end{table*}

\begin{figure*}[t]
\centering
    \subfigure[LaMP-4-trained LLM applied to LaMP-5.]
    {
    \includegraphics[width=0.32 \linewidth]{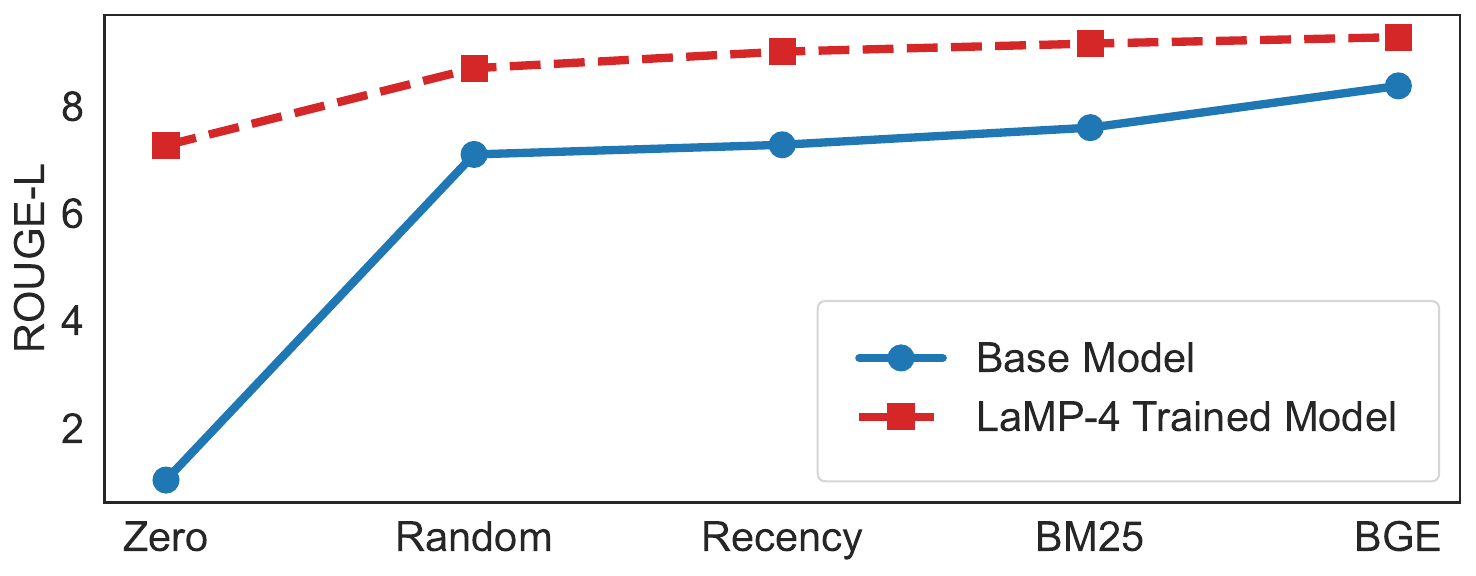}
    \label{fig:cross_4_for_5}
    }
    \subfigure[LaMP-5-trained LLM applied to LaMP-4..]
    {
    \includegraphics[width=0.32 \linewidth]{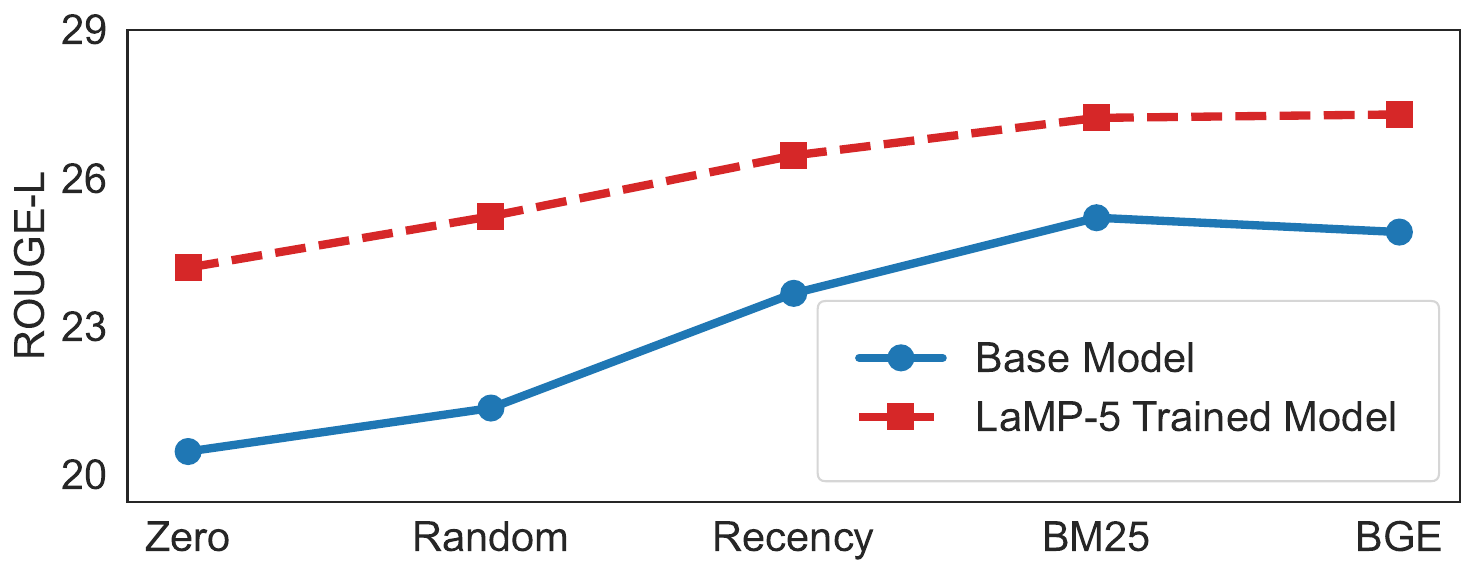}
    \label{fig:cross_5_for_4}
    }
    \subfigure[Different numbers of user
profiles in LaMP-5.]
    {
    \includegraphics[width=0.32 \linewidth]{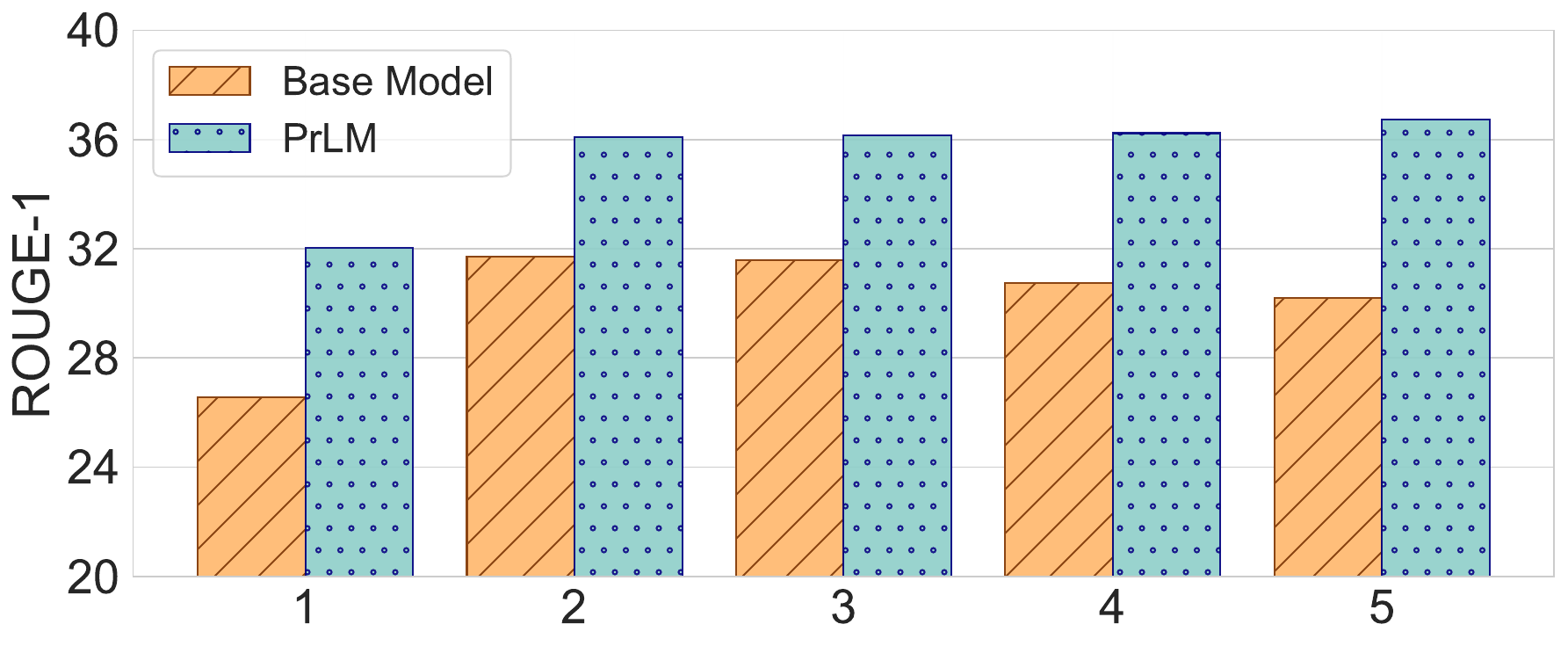}
    \label{fig:numbers}
    }
    \caption{The Base model is the original Deepseek-R1-Distill-Qwen1.5B. (a) and (b): Cross domain Results. The X-axis represents the retrieval strategy. (c) The X-axis represents the number of historical user behaviors retrieved. }
    \vspace{-5mm}
\end{figure*}

\subsection{Experimental Settings}
\subsubsection{Dataset and Metric}
We use the datasets from the LaMP~\cite{salemi2024lamp}, selecting LaMP-4: Personalized News Headline Generation; LaMP-5: Personalized Scholarly Title Generation; and LaMP-7: Personalized Tweet Paraphrasing. 
We use the temporal splitting.
Statistical analyses are in Table~\ref{tab:data_main}.
Following~\cite{shi2025retrieval}, we evaluate the outputs using ROUGE-1, ROUGE-2, ROUGE-L, and BLEU.

\subsubsection{Comparison Methods}
PrLM is retrieval-agnostic approach, we compared it with the following methods:
\textbf{Zero-Shot}: Directly input the user's query into the LLM to obtain the output without any retrieval. This is a non-personalized baseline.
\textbf{Random}: Randomly select k user profiles.
\textbf{Recency}: Select the most recent k user profiles according to the time.
\textbf{BM25}~\cite{robertson1995okapi}: Retrieve the top k user profiles using BM25 based on the user's query.
\textbf{BGE}~\cite{bge_embedding}: Retrieve the top k user profiles using BGE based on the user's query.
\textbf{ROPG}~\cite{salemi2024optimization}: Retrieval after fine-tuning BGE with feedback from the LLM.
\textbf{CFRAG}~\cite{shi2025retrieval}: Introducing similar user retrieval, retrieving first and then re-ranking.
We alse compare with \textbf{w/o $r_\mathrm{personal}$}.

\subsubsection{Implementation Details}
We retrieve five documents and use greedy decoding during generation to make the results easily reproducible.
BGE is bge-base-en-v1.5~\cite{bge_embedding}.
We trained the PrLM using GRPO. We sampled 4 outputs each time, with a maximum completion length of 768. We employed LoRA~\cite{hu2022lora} for efficient training, where the rank of LoRA was set to 16 and the alpha was also 16. The learning rate was set to 1e-6. In Eq.~\ref{eq:reward}, both $\alpha$ and $\beta$ were set to 0.1. 
We used DeepSpeed ZeRO-2~\cite{rajbhandari2020zero} and vLLM~\cite{kwon2023efficient} for training and testing.
For personalized reward model training, we trained for 3 epochs with a learning rate of 1e-6 and a maximum length of 512. We trained the BERT-base-uncased model.
We conducted our experiments on A6000 GPUs. More details can be found at \url{https://github.com/ke-01/PrLM}.

\subsection{Main Results}
The experimental results are shown in Table~\ref{tab:merged_results}. We observe that:

\textbf{User profiles are important.} It can be seen that as long as the user's historical behavior is retrieved and added to the prompt of the LLM, the output of the model is better than the zero-shot setting. This is similar to the conclusions of previous work~\cite{shi2025retrieval}. The user's historical behavior helps the LLM to better capture the user's preferences, thereby providing more personalized outputs. 

\textbf{PrLM achieves the best performance.} We can see that in all datasets, compared with all the baselines, PrLM achieved the best results. This indicates the effectiveness of PrLM. The personalized reward model helps guide the LLM to better capture the user's true preferences, thereby generating more personalized outputs.

\textbf{Both the retriever and the LLM are important.} We can see that the retriever is very important. 
For example, we can see that in LaMP-5, the effect of BGE retrieval is not as good as that of BM25. 
In addition, PrLM and w/o $r_{personal}$ trained the LLM, making the LLM perform better than the original LLM. This shows that training a better LLM is also very important for personalized RAG.

\subsection{Retriever Robustness}
In this section, we apply the LLM trained with documents retrieved by the BGE strategy to different retrieval strategies. 
The results are shown in Table~\ref{tab:retrieve}.
It can be seen that adding PrLM achieves improvements across all retrieval strategies. This indicates that PrLM has high robustness to different retrieval strategies. 
PrLM enables the LLM to better utilize the user profiles.

\subsection{Cross-domain Robustness}
In this section, we investigate how PrLM performs on cross-domain tasks. Specifically, we apply the model trained on the LaMP-4 to the LaMP-5 and vice versa. 
We conduct experiments under different retrieval strategies. 
The results are shown in Figure~\ref{fig:cross_4_for_5} and \ref{fig:cross_5_for_4}. It can be seen that when the model trained on the LaMP-4 is applied to LaMP-5, and vice versa, PrLM achieves better performance. This indicates that PrLM has good cross-domain capabilities. This further demonstrates the robustness of PrLM.

\subsection{Robustness of User Profiles Numbers}
In this section, we investigate the model's performance when retrieving different numbers of user profiles, with the results shown in Figure~\ref{fig:numbers}.
It can be observed that the performance of Deepseek-R1-Distill-Qwen-1.5B initially increases and then decreases as the number of retrieved user profiles grows. 
This is because a larger volume of user behaviors introduces more noise, making it difficult for the model to distinguish which parts of the user behaviors are relevant.
In contrast, PrLM's performance gradually improves with the increase in the number of documents and does not decline. This indicates that PrLM exhibits better robustness and generalizability, effectively utilizing the user profiles 

\section{Conclusion}
In this work, we propose \texttt{PrLM}, a reinforcement learning framework for personalized retrieval-augmented generation that enables large language models to perform explicit reasoning over retrieved user profiles. Specifically, we adopt a group-based reinforcement learning strategy to guide the model's reasoning process to address the lack of annotated reasoning supervision. Additionally, we introduce a contrastive reward modeling approach to assess the personalization quality of LLM-generated responses. Experimental results on three personalized generation benchmarks demonstrate that \texttt{PrLM} consistently improves response quality and enhances robustness across varying retrieval settings.

\section*{GenAI Usage Disclosure}
I know that the ACM’s Authorship Policy requires full disclosure of all use of generative AI tools in all stages of the research (including the code and data) and the writing.
No GenAI tools were used in any stage of the research, nor in the writing.

\bibliographystyle{ACM-Reference-Format}
\bibliography{sample-base}

\appendix

\end{document}